\pdfoutput=1
\documentclass[nofootinbib,superscriptaddress,amsfonts,amssymb,amsmath, twocolumn]{revtex4-1}
\usepackage{amsmath,bm,mathtools,physics}
\usepackage[utf8]{inputenc}
\numberwithin{equation}{section}

\usepackage{perpage} 
\MakePerPage{footnote}

\usepackage{graphicx}
\usepackage{comment}
\graphicspath{ {/} }
\newcommand{\ve}[1]{\bm{#1}}
\newcommand{\secref}[1]{section~#1}
\newcommand{\figref}[1]{figure~#1}
\newcommand{\equref}[1]{eqn.~(#1)}

\usepackage{tikz}
\usetikzlibrary{trees}
\usetikzlibrary{decorations.pathmorphing}
\usetikzlibrary{decorations.markings}
\usetikzlibrary{shapes.misc}

\tikzset{
	boundaryInsertion/.style={
		circle,
		draw,
		inner sep=1pt,
	},
	choose/.style={
		rounded corners=3,
		color=#1!50!black,
	},
	choose/.default=white,
	ball/.style={
		circle,
		draw,
		fill=black,
		inner sep=1pt,
		minimum size=1pt
	},
	line1/.style={
		thick,
		orange!80!black,
		postaction={
			decorate,
			decoration={
				markings,
				mark=at position 0.5 with {
					\draw (-2pt,-2pt) -- (2pt,2pt);
					\draw (2pt,-2pt) -- (-2pt,2pt);
				}
			}
		},
	},
	line2/.style={
		thick,
		red!50!black,
	},	
	gluon/.style={
		decoration={
			snake,
			amplitude=2.3pt,
			segment length=2pt,
		},
		decorate,
		blue!50!black,
	},
	threept/.style={
		circle,
		draw,
		inner sep=2pt,
	},
	twopt/.style={
		circle,
		draw,
		fill=black,
		inner sep=1pt,
		minimum size=1pt
	},
	cross/.style={
		cross out,
		draw=black, 
		minimum size=7pt, 
		inner sep=0pt,
		outer sep=0pt
	},
	scalar/.style={
		thick,
		dashed,
		postaction={
			decorate,
			decoration={
				markings,
				mark=at position 0.5 with {\arrow{>}}
			}
		}
	},
	spinning/.style={
		thick,
		postaction={
			decorate,
			decoration={
				markings,
				mark=at position 0.5 with {\arrow{>}}
			}
		}
	},
	spinning no arrow/.style={
		thick,
	},
	finite with arrow/.style={
		decoration={
			snake,
			amplitude=1pt,
			segment length=6pt,
			post length=2pt
		},
		decorate,
		thick,->
	},
	finite/.style={
		decoration={
			snake,
			amplitude=1pt,
			segment length=6pt,
		},
		decorate,
		thick
	}
}

\newcommand{\diagramEnvelope}[1]{#1}

\definecolor{darkblue}{rgb}{0.1,0.1,.7}
\usepackage[colorlinks, linkcolor=darkblue, citecolor=darkblue, urlcolor=darkblue, linktocpage]{hyperref} 

\begin{document}
\title{New relation for AdS amplitudes}
\author{Soner Albayrak}
\affiliation{Yale University, New Haven, CT 06511, USA}
\affiliation{Walter Burke Institute for Theoretical Physics, Caltech, Pasadena, CA 91125, USA}
\author{Chandramouli Chowdhury} 
\affiliation{International Centre for Theoretical Sciences, Tata Institute of Fundamental Research, Sivakote, Bangalore 560089, INDIA}
\author{Savan Kharel} 
\affiliation{Yale University, New Haven, CT 06511, USA}
\affiliation{Williams College, Williamstown, MA, 01267, USA}

\begin{abstract}
In this paper, we present a simple and iterative algorithm that computes Witten diagrams. We focus on the gauge correlators in AdS in four dimensions in momentum space. These new combinatorial relations will allow one to generate tree level amplitudes algebraically, without having to do any explicit bulk integrations; hence, leading to a simple method of calculating higher point gauge amplitudes.
	\end{abstract}
\date{\today}
\maketitle

\section{Introduction}
A great deal of developments have taken place in the last decade in the study of flat space scattering amplitudes of gauge theories. The modern amplitudes research program has led to many unexpected relations such as on shell recursion relations \cite{Britto:2005fq, Britto:2004ap}, the connection to mathematical structures like Grassmanian geometry, and the discovery of the amplituhedron \cite{ ArkaniHamed:2010gg,Arkani-Hamed:2013jha,Arkani-Hamed:2013kca}. For an introduction to these computational tools and an overview of these developments, we refer the reader to  \cite{Elvang:2013cua, Mangano:1990by,
Dixon:2013uaa,Henn:2014yza,Weinzierl:2016bus}.

Likewise, outstanding progress has been made in our understanding of quantum gravity with the discovery of the holographic principle \cite{Susskind:1994vu,tHooft:1993dmi}. The holographic principle implies that degrees of freedom that are encoded in the boundary in $d$ dimensions can describe the $d+1$ dimension interior of the spacetime. A concrete example of holography is the gauge/gravity duality, i.e. the correspondence between Anti-de Sitter space (AdS) with Conformal Field Theory (CFT) \cite{Maldacena:1997re, Witten:1998qj}. By relating the boundary operators to the bulk fields, the CFT correlation functions can be interpreted as AdS scattering amplitudes, and vice versa \cite{Penedones:2010ue,Heemskerk:2009pn}.

Despite the splendid advancements in these fields, the amplitude programs in flat space and AdS have remained relatively isolated\footnote{For some counter examples, see \cite{Rastelli:2016nze,Rastelli:2019gtj,Giusto:2019pxc}.} partly due to the the difficulty of position space calculations in AdS. Various complementary approaches, e.g. Mellin space, have been investigated to address these notorious computations \cite{Freedman:1998tz,DHoker:1999mqo,DHoker:1999kzh,Penedones:2010ue, Paulos:2011ie,  Mack:2009gy, Fitzpatrick:2011ia,Kharel:2013mka, Fitzpatrick:2011hu, Fitzpatrick:2011dm, Costa:2014kfa, Raju:2011mp,Raju:2012zr,Jepsen:2018ajn,Jepsen:2018dqp, Gubser:2018cha, Hijano:2015zsa, Yuan:2018qva}. We believe that the power of the momentum space perturbation theory has not been fully realized. Hence, we propose a new method to explicitly compute AdS$_4$ momentum space amplitudes.\footnote{By holography, our method can also be employed to compute dual CFT$_3$ correlators. We refer the reader to \cite{Bzowski:2015yxv, Bzowski:2011ab, Bzowski:2012ih, Bzowski:2013sza, Farrow:2018yni} for other momentum space approaches to conformal field theories.}

In this paper, we will outline a new diagrammatic method that will enable us to systematically compute higher point amplitudes. Crucially, this method bypasses the cumbersome bulk integrals, therefore reducing the computations of the correlators to simple algebraic relations. This framework can be utilized to calculate higher point tree level AdS amplitudes, which can be used as data points to extract physical and mathematical insights. We liken this to the similar methodology employed in flat space scattering amplitudes over the last decade, where explicit flat space amplitudes were used as data points to generate surprising relations, such as the BCJ duality, CHY relations, and the amplituhedron \cite{Bern:2008qj, Cachazo:2013hca, Cachazo:2013iea,  Arkani-Hamed:2013jha}.

Here is the brief outline of this paper. We begin, in \secref{\ref{sec:preliminaries}}, with a review of the formalism for momentum space gauge theory correlators. Then, we introduce bulk to bulk and bulk to boundary propagators for gauge fields as solutions to their respective equations of motion. 
In \secref{\ref{sec:algorithm}}, we present the combinatorial rule to compute \emph{part} of the scattering amplitudes for gauge fields in AdS. Subsequently in \secref{\ref{sec: getting crossed}}, we  discuss computation of the  \emph{remaining part}, hence obtaining the full expression. Finally, we summarize  and discuss promising future directions in \secref{\ref{sec:conclusion}}. We provide an appendix to further illustrate the main points of this paper with an additional example.

\section{Gauge fields in AdS}
\label{sec:preliminaries}
We are interested in a non-Abelian gauge theory in AdS, described by an action \mbox{$S\sim\int d^dx z^{-d-1}dz  F_{\mu \nu}^a F^{\mu \nu, a}$} where $z$ is the radial coordinate and $x_i$ approaches to the boundary coordinate as $z\rightarrow 0$ for the AdS metric \mbox{$ds^2 =z^{-2}(dz^2 +  \eta_{ij} dx^i dx^j)$} in the Poincar\'e patch. These coordinates make the Poincar\'e invariance manifest; thus, it is easy to transform the position space coordinates $x_i$ to the momentum space variables $k_i$. In this paper, we will focus on $d=3$, impose axial gauge, $A_0^a=0$; and work with the coordinates $\{z,k_i\}$, following closely the treatment of \cite{Raju:2010by}.

One can go ahead and solve the equations of motion: the normalizability at the boundary and the regularity in the bulk unambiguously determines the bulk to boundary propagator, i.e. \mbox{$A_i^a(z,\bm{k})=\ve{\epsilon}_i ^a \sqrt{z}  E_{1/2} ( k z)$} \cite{Raju:2011mp}. Here $\ve{\epsilon}$ is the transverse polarization vector; $k$ is the positive norm of the momentum $\bm{k}$, i.e. $k=\sqrt{\abs{\bm{k}^2}}$; and $E$ is a Bessel kind function, appropriate to the sign of $\bm{k}^2$.\footnote{
	In this paper, we will work with transition amplitudes, similar to \cite{Albayrak:2018tam,Albayrak:2019yve,Albayrak:2020isk}, and use $\ve{\epsilon}_i^a \sqrt{\frac{2 k z}{\pi}} K_{1/2}(k z)$ as our effective bulk-to-boundary propagator, same as \cite{Raju:2012zs,Albayrak:2018tam}.
}

\begin{figure}
	\centering
	\includegraphics[width=4.5cm]{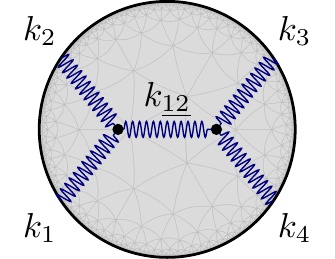}
	\caption{\label{fig:4pt}A four point Witten diagram}
\end{figure}

In this framework, the amplitude for the four point exchange diagram depicted in \figref{\ref{fig:4pt}} takes the form
\begin{equation*}
\begin{aligned}
\mathcal{M}_{4s}^{ijkl} =\frac{4\sqrt{k_1k_2k_3k_4}}{i\pi^2}
V^{ijm}(\ve{k}_1, \ve{k}_2, -\ve{k}_{12} )  V^{kln} (\ve{k}_3, \ve{k}_4, \ve{k}_{12}) \\\times
\int \omega\, d\omega\, d z\, d z'
K_{1/2}(k_1 z) K_{1/2}(k_2 z)J_{1/2}(\omega z) K_{1/2}(k_3 z')
\\\times K_{1/2}(k_4 z')J_{1/2}(\omega z')
\frac{\eta_{mn}\omega^2+(\ve{k}_{12})_m(\ve{k}_{12})_n}{\omega^2( k_{12}^2 + \omega^2 )}\;,
\end{aligned}
\end{equation*}
which is obtained by gluing bulk to bulk and bulk to boundary propagators. In this expression, and in the rest of the paper, we suppress the color dependence for notational brevity.

In the above expression, we use 
\begin{equation*}
V_{ijk}(\ve{k}_i)\coloneqq{}
\frac{\eta_{ij}(\ve{k}_1-\ve{k}_2)_k+\eta_{jk}(\ve{k}_2-\ve{k}_3)_i+\eta_{ki}(\ve{k}_3-\ve{k}_1)_j}{-i\sqrt{2}}
\end{equation*}
which is the three point vertex for color-ordered amplitudes \cite{Dixon:1996wi}. For later convenience, we also define the four point vertex factor
\begin{equation*}
V^{ijkl}_c\coloneqq{}{} i \;\eta^{ik} \eta^{jl}-\frac{i}{2}\left(\eta^{ij} \eta^{kl}+\eta^{il} \eta^{jk}\right)\;.
\end{equation*}
In order to make the notation concise, we use \mbox{$V_{12k}\equiv\ve{\epsilon}_1^i\ve{\epsilon}_2^jV_{ijk}$} and likewise for other tensors. As we will not be working with individual components, such notation does not lead to any ambiguity.

One can now calculate the full amplitude by directly carrying out the bulk integrals, thus arriving at the four point expression:\footnote{In this paper, we use the notation of \cite{Albayrak:2018tam} to denote sums of momenta, i.e.
\begin{subequations}
\begin{multline}
k_{\underline{i_{11}i_{12}\dots i_{1n_1}}\;\underline{i_{21}i_{22}\dots i_{2n_2}}\dots \underline{i_{m1}i_{m2}\dots i_{mn_m}}j_1j_2\dots j_p}\\\coloneqq\sum\limits_{a=1}^{m}\abs{\sum\limits_{b=1}^{n_a}\bm{k}_{i_{ab}}}+\sum\limits_{c=1}^{p}\abs{\bm{k}_{j_c}}\;,
\end{multline}
and
\begin{equation}
\bm{k}_{i_1i_2\dots i_n}\coloneqq\bm{k}_{i_1}+\bm{k}_{i_2}+\cdots+\bm{k}_{i_n}\;.
\end{equation}
E.g., $k_{\underline{12}3\underline{45}}\equiv\abs{\bm{k}_1+\bm{k}_2}+\abs{\bm{k}_3}+\abs{\bm{k}_4+\bm{k}_5}\;$ and $\bm{k}_{12}\equiv\bm{k}_1+\bm{k}_2$.
\end{subequations}
}
\begin{multline}
\mathcal{M}_{4s} =-i\frac{ V^{12m}(\ve{k}_{1}, \ve{k}_{2},-\ve{k}_{{12}})  V^{34n} (\ve{k}_{3}, \ve{k}_{4}, \ve{k}_{{12}} ) }{k_{1234} k_{12\underline{12}} k_{34\underline{12}}}
\\\times
\left(\eta_{mn}+\frac{k_{1234 \underline{12}}\left(\ve{k}_{12}\right)_m\left(\ve{k}_{12}\right)_n}{k_{12} k_{34}  k_{\underline{12}}}\right)
\label{eq: 4s contribution}\;.
\end{multline}
Note that this is the amplitude associated with the $s-$channel diagram. Using the same method we can compute the $t-$channel and contact diagrams.

The form of the expression above suggests that we can decompose any tree-level diagram into two parts: the vertex factors carrying the dependence on the individual vectors $\bm{k}_i$, and the rest of the amplitude that we will call \emph{the truncated diagram}. For example, in \equref{\ref{eq: 4s contribution}},
we see that $k_1$ dependence enters into the truncated piece only through the terms $k_{12}$ and $k_{\underline{12}}$. Note that apparently different pieces are combinations of such terms, e.g. $k_{1234\underline{12}}=k_{12}+k_{\underline{12}}+k_{34}$.

We can further decompose the truncated diagram into two distinct scalar diagrams, \emph{straight} and \emph{crossed},
\begin{equation}
\label{eq: propagator decomposition}
\hspace{-.03in}
\begin{aligned}
\diagramEnvelope{\begin{tikzpicture}[anchor=base,baseline]
	\node (left) at (-.5,0) [ball] {};
	\node (right) at (.5,0) [ball] {};
	\node at (-.5,0) [below] {$p_1$};
	\node at (-.5,0) [above] {$m$};
	\node at (.5,0) [below] {$p_2$};
	\node at (.5,0) [above] {$n$};
	\node at (0,0) [above] {$k$};
	\draw[gluon] (left)-- (right);
	\end{tikzpicture}}
\end{aligned}
=
\Pi^{(1)mn}_{k}
\begin{aligned}
\diagramEnvelope{\begin{tikzpicture}[anchor=base,baseline]
	\node (left) at (-.5,0) [ball] {};
	\node (right) at (.5,0) [ball] {};
	\node at (-.5,0) [below] {$p_1$};
	\node at (.5,0) [below] {$p_2$};
	\node at (0,0) [above] {$k$};
	\draw[line2] (left)-- (right);
	\end{tikzpicture}}
\end{aligned}+
\Pi^{(2)mn}_{k}
\begin{aligned}
\diagramEnvelope{\begin{tikzpicture}[anchor=base,baseline]
	\node (left) at (-.5,0) [ball] {};
	\node (right) at (.5,0) [ball] {};
	\node at (-.5,0) [below] {$p_1$};
	\node at (.5,0) [below] {$p_2$};
	\node at (0,0) [above] {$k$};
	\draw[line1] (left)-- (right);
	\end{tikzpicture}}
\end{aligned}
\end{equation}
where we define the projectors
\begin{equation}
\label{eq: projectors}
\Pi^{(1)mn}_k\equiv{}\frac{\eta_{mn}k^2-\bm{k}_m\bm{k}_n}{ik^2}\;,\quad 
\Pi^{(2)mn}_k\equiv{}{}\frac{\bm{k}_m\bm{k}_n}{ik^2}\;.
\end{equation}

\begin{figure}
	\centering
	\begin{equation*}
\begin{aligned}
\includegraphics[width=4cm]{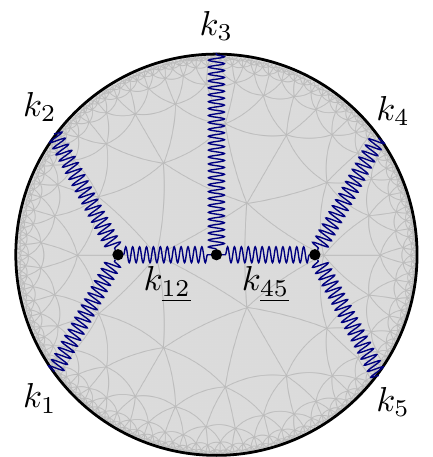}
\end{aligned}\Huge\Rightarrow
\begin{aligned}
\includegraphics[scale=.8]{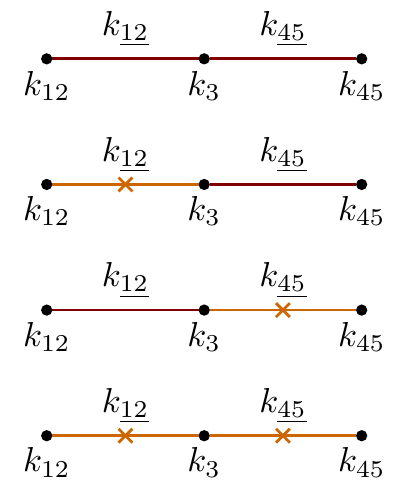}
\end{aligned}
	\end{equation*}
\caption{\label{fig:5ptdecomposition}
Decomposition of a five point Witten diagram into scalar graphs}
\end{figure}

To be specific, $\begin{aligned}
\diagramEnvelope{\begin{tikzpicture}[anchor=base,baseline]
	\node (left) at (-.5,0) [ball] {};
	\node (right) at (.5,0) [ball] {};
	\node at (-.5,0) [below] {$p_1$};
	\node at (.5,0) [below] {$p_2$};
	\node at (-.5,0) [above] {$m$};
	\node at (.5,0) [above] {$n$};
	\node at (0,0) [above] {$k$};
	\draw[gluon] (left)-- (right);
	\end{tikzpicture}}
\end{aligned}$  denotes the bulk point integrated-propagator in the momentum space, which coincides with the truncated four point amplitude. We will use this graph as a basis element to construct higher point truncated diagrams: working with these bulk-point integrated diagrams will allows us to efficiently extract several different Witten diagrams from their common truncated graph. For example, we can connect two of these basis elements to construct the amputated graph necessary for the five point Witten diagram, as can be seen in \figref{\ref{fig:5ptdecomposition}}.

The advantage of the decomposition in \equref{\ref{eq: propagator decomposition}} is the simplicity of the scalar graphs: they are fully agnostic to what is attached at the vertices as long as we know the sum of norms of the momenta that flows into that vertex. This simply follows from the form of our effective bulk to boundary propagators as their bulk-point dependencies at the vertices are merely additive. For example,
$p_{1,2}$ in \equref{\ref{eq: propagator decomposition}} represents the sum of the norms of bulk to boundary momenta, i.e. $p_1=k_1+k_2+\dots ~$. 

This framework suggests that any tree-level Witten diagram can be decomposed into sums and products of vertex factors, projectors, and several scalar graphs; thus, the calculation of a Witten diagram reduces to a calculation of scalar graphs. E.g., for the five point diagram given in \figref{\ref{fig:5ptdecomposition}}, we can obtain the full expression once we compute the corresponding scalar graphs.

The calculations of graphs with crossed lines and graphs without crossed lines are different: \emph{we will deal with them separately}. We will present the algorithm to compute the graphs of the straight lines in the next section: this algorithm exists in the literature albeit in a completely different context and theory \cite{Benincasa:2018ssx, Arkani-Hamed:2018bjr, Arkani-Hamed:2017fdk, BenincasaDraft}.  After that, we will establish the connection between the crossed and straight lines in
\secref{\ref{sec: getting crossed}}. Such a relation will enable us to compute the complete amplitude.

\section{An algorithm to compute scalar graphs}
\label{sec:algorithm}
In \equref{\ref{eq: propagator decomposition}}, we showed that the gauge propagator can be decomposed into two parts. In this section, we will focus on the graphs of only straight lines. This is because such graphs satisfy a nice algorithmic relation that we will discuss below. We can understand this if we examine the explicit expression corresponding to the straight line:
\begin{equation*}
\begin{aligned}
\diagramEnvelope{\begin{tikzpicture}[anchor=base,baseline]
	\node (left) at (-.5,0) [ball] {};
	\node (right) at (.5,0) [ball] {};
	\node at (-.5,0) [below] {$p_1$};
	\node at (.5,0) [below] {$p_2$};
	\node at (0,0) [above] {$k$};
	\draw[line2] (left)-- (right);
	\end{tikzpicture}}
\end{aligned}
=\int\limits_{0}^{\infty}\frac{dz_1dz_2}{(z_1z_2)^{4}}G_s(k,z_1,z_2) B(z_1,p_1;z_2,p_2)
\end{equation*}
for
\begin{equation}
\label{eq: straight line expression}
G_s(k,z_1,z_2)\equiv \int\limits_{0}^{\infty}
\frac{\omega d\omega \sqrt{z_1 z_2} J_{1/2}(\omega z_1) J_{1/2}(\omega z_2)}{ k^2+\omega^2-i \epsilon}\;,
\end{equation}
where $B(z_1,p_1;z_2,p_2)$ encodes the contribution of other graphs that connect to this propagator at the bulk points $z_1$ and $z_2$. 
As we discussed above, these contributions are additive and are represented by the letter $p$ in the diagram.

The remarkable feature of $G_s$ is that it is proportional to the \emph{cosmological} propagator derived for the conformally coupled scalar; specifically
\begin{equation}
G_s(k,z,z')
\rightarrow\frac{i}{2} G_e(E_e,\eta_{\nu_e},\eta_{\nu_e'})
\end{equation}
where $\{z,k\}\rightarrow\{-\eta_{\nu_e},-i E_e\}$ in the notation of \cite{Arkani-Hamed:2017fdk}. In that paper, the authors show that one can compute similar graphs using algebraic means. The nice feature of our bulk-point integrations is that they do not extirpate this formalism, hence we can also use graph-wise calculations in our setting.\footnote{The proportionality factor of $i/2$ is necessary between the propagators to identify the graphs: the $\eta-$integration range is effectively half of $z-$integration and $i$ accounts for $k\rightarrow -iE_e$ in the graphs.} Here, we summarize our prescription to calculate any tree-level Witten diagram for AdS$_4$ gauge bosons in momentum space in axial gauge:
\begin{enumerate}
	\item Draw the relevant Witten diagram and truncate it
	\item Decompose the truncated diagram in terms of straight and crossed lines 
	\item There is always a unique \emph{straight-only} scalar graph which does not have any crossed lines. Calculate that graph by mere algebraic means
	\item Obtain the other scalar graphs by the procedure described in \secref{\ref{sec: getting crossed}}
	\item Combine all scalar graphs with the relevant projectors and the vertex factors to obtain the full Witten diagram
\end{enumerate}

We have already explained the first two items, let us now move on to the third point. For that, we need to review the algorithm of \cite{Arkani-Hamed:2017fdk} which we will use to procure expressions for straight-only scalar graphs.

One starts with the full diagram and  considers the ways to decompose it into different subdiagrams by \emph{cutting} the lines. Then, one associates an expression to each decomposition and sums these partial amplitudes.

The partial amplitude for a particular decomposition is simply the product of the expressions associated with the subgraphs. For a particular subgraph, the associated expression is inverse of the sum of all \emph{vertex norms} within that subgraph and \emph{line norms} going out of that subgraph. With this rule, we associate the corresponding expressions to all subgraphs, starting from the full graph itself.

\begin{figure}
	\centering
	\begin{tabular}{l}
		\hskip .7in $\includegraphics[scale=1]{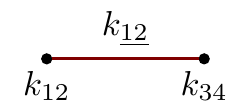}$
		\\
		\hskip 1.15in
		$\Downarrow$
		\\
		$\includegraphics[scale=1]{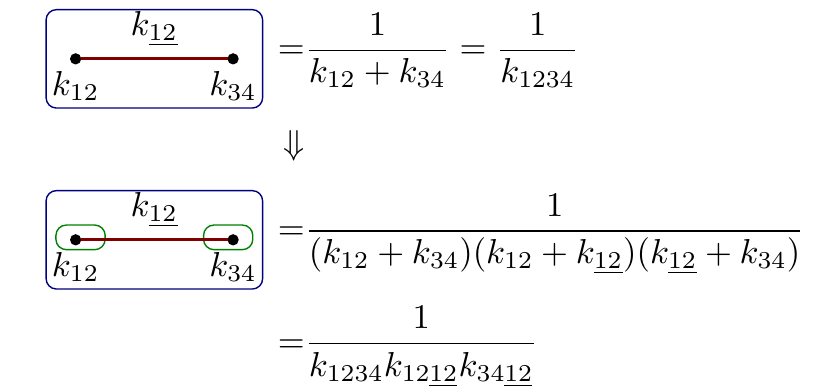}$
	\end{tabular}
	\caption{Algorithm to compute the amplitudes of straight lines\label{fig:algorithm}}
\end{figure}

Let us clarify this rather formal explanation with an explicit example: a single straight line, depicted in \figref{\ref{fig:algorithm}}. As we see, there is only one possible decomposition because there is only one line to cut! For this partial-amplitude, there are three subgraphs: the full graph indicated by the blue rectangle and its subgraphs indicated by the green rectangles. Since there is no line crossing the blue rectangle, its corresponding expression is inverse of the sum of the vertex norms, i.e. $\left(k_{12}+k_{34}\right)^{-1}=k_{1234}^{-1}$. On the contrary, the green rectangles get contributions from the line norm as well, hence they are $k_{12\underline{12}}^{-1}$ and $k_{34\underline{12}}^{-1}$. The full partial amplitude is simply the product of these three terms: \mbox{$\mathcal{A}^{(1)}=\left(k_{1234} k_{12\underline{12}} k_{34\underline{12}}\right)^{-1}$}.

The expression of the blue rectangle satisfies the generic feature of our algorithm: since there is always a graph than encapsulates the full diagram,
all amplitudes will have a factor in the denominator, which is the sum of all vertex norms, i.e. $k_{1}+k_2+\cdots +k_n$. As this factor will always be multiplicative, our algorithm guarantees that the amplitudes for \emph{straight-only} scalar graphs will have a pole where $k_{123\dots n}\rightarrow 0$ for any tree-level $n-$point Witten diagram. 

The appearance of such poles can be understood in the context of flat-space limit due to a nifty relation:
\begin{equation}
\mathcal{M}_{\text{flat-space}}=\operatorname*{Res}_{k_{123\dots n}\rightarrow 0}\mathcal{M}\;.
\end{equation}
This simply follows from our choice of momenta variables and the momentum conservation condition in flat space \cite{Maldacena:2011nz,Raju:2012zr,Arkani-Hamed:2017fdk}; hence flat space limits of our expressions are immediately manifest.

A more non-trivial example is the graph with two straight lines, which is relevant for five point Witten diagrams. In this case we have two lines to cut, indicating two distinct decompositions. We calculate the partial amplitude for each decomposition and take their sum:
\begin{equation}
\footnotesize
\begin{aligned}
{}&\diagramEnvelope{\begin{tikzpicture}[anchor=base,baseline]
	\node (left) at (-.8,0) [ball] {};
	\node (middle) at (.8,0) [ball] {};
	\node (right) at (2.4,0) [ball] {};
	\node at (-.8,0) [below] {$k_{12}$};
	\node at (.8,0) [below] {$k_{3}$};
	\node at (2.4,0) [below] {$k_{45}$};
	\node at (0,0) [above] {$k_{\underline{12}}$};
	\node at (1.6,0) [above] {$k_{\underline{45}}$};
	\draw[line2] (left)-- (middle);
	\draw[line2] (right)-- (middle);
	\end{tikzpicture}}
=
\\
&\left(
\diagramEnvelope{\begin{tikzpicture}[anchor=base,baseline]
	\draw[choose=blue] (-1.2,-.6) rectangle (2.7cm,.6cm);
	\draw[choose=green] (-1.1,-.5) rectangle (1cm,.5cm);
	\draw[choose=green] (2.1,-.5) rectangle (2.65cm,.5cm);
	\draw[choose=magenta] (-1,-.1) rectangle (-.4cm,.15cm);
	\draw[choose=magenta] (.35,-.1) rectangle (.95cm,.15cm);
	\node (left) at (-.8,0) [ball] {};
	\node (middle) at (.8,0) [ball] {};
	\node (right) at (2.4,0) [ball] {};
	\node at (-.8,0) [below] {$k_{12}$};
	\node at (.8,0) [below] {$k_{3}$};
	\node at (2.4,0) [below] {$k_{45}$};
	\node at (0,0) [above] {$k_{\underline{12}}$};
	\node at (1.6,0) [above] {$k_{\underline{45}}$};
	\draw[line2] (left)-- (middle);
	\draw[line2] (right)-- (middle);
	\end{tikzpicture}}
=\frac{1}{k_{12345}k_{123\underline{45}}k_{45\underline{45}}k_{12\underline{12}}k_{\underline{12}3\underline{45}}}
\right)
\\&+
\left(
\diagramEnvelope{\begin{tikzpicture}[anchor=base,baseline]
	\draw[choose=blue] (-1.2,-.6) rectangle (2.7cm,.6cm);
	\draw[choose=green] (-1.1,-.5) rectangle (-.5cm,.5cm);
	\draw[choose=green] (0.6,-.5) rectangle (2.65cm,.5cm);
	\draw[choose=magenta] (0.65,-.1) rectangle (1.25cm,.15cm);
	\draw[choose=magenta] (2,-.1) rectangle (2.6cm,.15cm);
	\node (left) at (-.8,0) [ball] {};
	\node (middle) at (.8,0) [ball] {};
	\node (right) at (2.4,0) [ball] {};
	\node at (-.8,0) [below] {$k_{12}$};
	\node at (.8,0) [below] {$k_{3}$};
	\node at (2.4,0) [below] {$k_{45}$};
	\node at (0,0) [above] {$k_{\underline{12}}$};
	\node at (1.6,0) [above] {$k_{\underline{45}}$};
	\draw[line2] (left)-- (middle);
	\draw[line2] (right)-- (middle);
	\end{tikzpicture}}
=\frac{1}{k_{12345}k_{12\underline{12}}k_{\underline{12}345}k_{\underline{12}3\underline{45}}k_{45\underline{45}}}
\right)
\end{aligned}
\end{equation}

which yields a surprisingly compact final amplitude
\begin{equation}
\label{eq: five point straight amplitude}
\mathcal{A}^{(11)}=\frac{k_{12 \underline{12}3345\underline{45}}}{k_{12345} k_{12 \underline{12}} k_{345 \underline{12}} k_{3 \underline{12}\;\underline{45}} k_{45
		\underline{45}} k_{123 \underline{45}}}\;.
\end{equation}
This is the same expression computed by brute force calculation in \cite{Albayrak:2018tam}.

\section{From straight graph to crossed graph}
\label{sec: getting crossed}

We have presented an elegant formalism in the previous section, and showed how one can easily extract the amplitude for a straight-only graph. An astute reader may object that despite its efficiency this formalism only yields a small part of the full amplitude. For instance, we need to calculate three more graphs if we want to obtain the amplitude for a five point Witten diagram, as seen in \figref{\ref{fig:5ptdecomposition}}. The situation appears to deteriorate at higher point computations as there are exponentially more graphs to calculate.

Below, we will demonstrate the opposite: all graphs are actually tied to the straight-only scalar graph hence one does not need to explicitly compute them once the straight-only scalar graph is obtained. This follows from the form of the gauge propagator in our settings:
\begin{flalign}
\mathcal{G}_{ij}(\ve{k}; z, z')=
\left(\eta_{ij}-\frac{\ve{k}_i\ve{k}_j}{k^2}\right)
\int
\omega d\omega 
\frac{\sqrt{z z'} J_{\frac{1}{2}}(\omega z) J_{{\frac{1}{2}}}(\omega z')}{i(k^2+\omega^2-i \epsilon)}
\nonumber\\
\hskip -10in
+
\frac{\ve{k}_i\ve{k}_j}{k^2}
\int
d\omega
\frac{k^2+\omega^2}{\omega} 
\frac{\sqrt{z z'} J_{\frac{1}{2}}(\omega z) J_{{\frac{1}{2}}}(\omega z')}{i(k^2+\omega^2-i \epsilon)}
\hspace{.4in}
\end{flalign}

This split form of the propagator is exactly what motivated us to introduce the decomposition in \equref{\ref{eq: propagator decomposition}} in the first place. Also, this reveals that the straight and crossed lines are related in a beautiful way:
\begin{equation}
\begin{aligned}
\diagramEnvelope{\begin{tikzpicture}[anchor=base,baseline]
	\node (left) at (-.8,0) [ball] {};
	\node (right) at (.8,0) [ball] {};
	\node at (-.8,0) [below] {$p_1$};
	\node at (.8,0) [below] {$p_2$};
	\node at (0,0) [above] {$k$};
	\draw[line1] (left)-- (right);
	\end{tikzpicture}}
\end{aligned}
=\lim\limits_{k\rightarrow 0}
\begin{aligned}
\diagramEnvelope{\begin{tikzpicture}[anchor=base,baseline]
	\node (left) at (-.8,0) [ball] {};
	\node (right) at (.8,0) [ball] {};
	\node at (-.8,0) [below] {$p_1$};
	\node at (.8,0) [below] {$p_2$};
	\node at (0,0) [above] {$k$};
	\draw[line2] (left)-- (right);
	\end{tikzpicture}}
\end{aligned}
\end{equation}
This simple relation of straight and crossed lines explains why the apparent problem of exponential increase in the number of total graphs poses no issue in the actual calculations: one can simply write the full expression as a simple operator acting on the straight-only graph: 
\begin{equation*}
\mathcal{M}^{mn}\equiv
\hspace{-0.1in}
\begin{aligned}
\diagramEnvelope{\begin{tikzpicture}[anchor=base,baseline]
	\node (left) at (-.5,0) [ball] {};
	\node (right) at (.5,0) [ball] {};
	\node at (-.5,0) [below] {$k_{12}$};
	\node at (.5,0) [below] {$k_{34}$};
	\node at (-.5,0) [above] {$m$};
	\node at (.5,0) [above] {$n$};
	\node at (0,0) [above] {$k_{\underline{12}}$};
	\draw[gluon] (left)-- (right);
	\end{tikzpicture}}
\end{aligned}
\hspace{-0.1in}
=\left(
\Pi^{(1)mn}_{k_{\underline{12}}}
+
\Pi^{(2)mn}_{k_{\underline{12}}}\lim\limits_{k_{\underline{12}}\rightarrow 0}
\right)
\hspace{-0.1in}
\begin{aligned}
\diagramEnvelope{\begin{tikzpicture}[anchor=base,baseline]
	\node (left) at (-.5,0) [ball] {};
	\node (right) at (.5,0) [ball] {};
	\node at (-.5,0) [below] {$k_{12}$};
	\node at (.5,0) [below] {$k_{34}$};
	\node at (0,0) [above] {$k_{\underline{12}}$};
	\draw[line2] (left)-- (right);
	\end{tikzpicture}}
\end{aligned}.
\end{equation*}

Alternatively, first we can find the amplitude for each graph and then combine them. In case of \figref{\ref{fig:5ptdecomposition}}, we write
\begin{subequations}
\label{eq: five point amplitudes}
\begin{align}
\mathcal{A}^{(12)}\equiv\begin{aligned}
\diagramEnvelope{\begin{tikzpicture}[anchor=base,baseline]
	\node (left) at (-.8,0) [ball] {};
	\node (middle) at (.8,0) [ball] {};
	\node (right) at (2.4,0) [ball] {};
	\node at (-.8,0) [below] {$k_{12}$};
	\node at (.8,0) [below] {$k_{3}$};
	\node at (2.4,0) [below] {$k_{45}$};
	\node at (0,0) [above] {$k_{\underline{12}}$};
	\node at (1.6,0) [above] {$k_{\underline{45}}$};
	\draw[line2] (left)-- (middle);
	\draw[line1] (right)-- (middle);
	\end{tikzpicture}}
\end{aligned}=&\lim\limits_{k_{\underline{45}}\rightarrow 0}\mathcal{A}^{(11)}
\\
\mathcal{A}^{(21)}\equiv\begin{aligned}
\diagramEnvelope{\begin{tikzpicture}[anchor=base,baseline]
	\node (left) at (-.8,0) [ball] {};
	\node (middle) at (.8,0) [ball] {};
	\node (right) at (2.4,0) [ball] {};
	\node at (-.8,0) [below] {$k_{12}$};
	\node at (.8,0) [below] {$k_{3}$};
	\node at (2.4,0) [below] {$k_{45}$};
	\node at (0,0) [above] {$k_{\underline{12}}$};
	\node at (1.6,0) [above] {$k_{\underline{45}}$};
	\draw[line1] (left)-- (middle);
	\draw[line2] (right)-- (middle);
	\end{tikzpicture}}
\end{aligned}=&\lim\limits_{k_{\underline{12}}\rightarrow 0}\mathcal{A}^{(11)}
\\
\mathcal{A}^{(22)}\equiv\begin{aligned}
\diagramEnvelope{\begin{tikzpicture}[anchor=base,baseline]
	\node (left) at (-.8,0) [ball] {};
	\node (middle) at (.8,0) [ball] {};
	\node (right) at (2.4,0) [ball] {};
	\node at (-.8,0) [below] {$k_{12}$};
	\node at (.8,0) [below] {$k_{3}$};
	\node at (2.4,0) [below] {$k_{45}$};
	\node at (0,0) [above] {$k_{\underline{12}}$};
	\node at (1.6,0) [above] {$k_{\underline{45}}$};
	\draw[line1] (left)-- (middle);
	\draw[line1] (right)-- (middle);
	\end{tikzpicture}}
\end{aligned}=&\lim\limits_{\substack{k_{\underline{12}}\rightarrow 0\\
		k_{\underline{45}}\rightarrow 0}}\mathcal{A}^{(11)}
\end{align}
\end{subequations}
for
\begin{equation}
\mathcal{A}^{(11)}\equiv\begin{aligned}
\diagramEnvelope{\begin{tikzpicture}[anchor=base,baseline]
	\node (left) at (-.8,0) [ball] {};
	\node (middle) at (.8,0) [ball] {};
	\node (right) at (2.4,0) [ball] {};
	\node at (-.8,0) [below] {$k_{12}$};
	\node at (.8,0) [below] {$k_{3}$};
	\node at (2.4,0) [below] {$k_{45}$};
	\node at (0,0) [above] {$k_{\underline{12}}$};
	\node at (1.6,0) [above] {$k_{\underline{45}}$};
	\draw[line2] (left)-- (middle);
	\draw[line2] (right)-- (middle);
	\end{tikzpicture}}
\end{aligned}
\end{equation}
whose explicit expression is given in \eqref{eq: five point straight amplitude}.

Incorporating all $\mathcal{A}^{(ij)}$ with the projectors, we get the full two line truncated diagram:
\begin{flalign}
\label{eq: decomposition of 5 pt truncated amplitude}
\hskip -.1in
\mathcal{M}^{mnpr}\equiv&\begin{aligned}
\diagramEnvelope{\begin{tikzpicture}[anchor=base,baseline]
	\node (left) at (-.8,0) [ball] {};
	\node (middle) at (.8,0) [ball] {};
	\node (right) at (2.4,0) [ball] {};
	\node at (-.8,0) [below] {$k_{12}$};
	\node at (.8,0) [below] {$k_{3}$};
	\node at (2.4,0) [below] {$k_{45}$};
	\node at (0,0) [above] {$k_{\underline{12}}$};
	\node at (1.6,0) [above] {$k_{\underline{45}}$};
	\node at (-.8,0) [above] {$m$};
	\node at (.65,0) [above] {$n$};
	\node at (.95,0) [above] {$p$};
	\node at (2.4,0) [above] {$r$};
	\draw[gluon] (left)-- (middle);
	\draw[gluon] (right)-- (middle);
	\end{tikzpicture}}
\end{aligned}
\nonumber\\
=&
\Pi^{(1)mn}_{k_{\underline{12}}}
\Pi^{(1)pr}_{k_{\underline{45}}}
\mathcal{A}^{(11)}
+
\Pi^{(1)mn}_{k_{\underline{12}}}
\Pi^{(2)pr}_{k_{\underline{45}}}
\mathcal{A}^{(12)}
\nonumber\\
&
\hskip -.05in
+
\Pi^{(2)mn}_{k_{\underline{12}}}
\Pi^{(1)pr}_{k_{\underline{45}}}
\mathcal{A}^{(21)}
+
\Pi^{(2)mn}_{k_{\underline{12}}}
\Pi^{(2)pr}_{k_{\underline{45}}}
\mathcal{A}^{(22)}
\end{flalign}

One should keep in mind that $\mathcal{M}^{mn}$ and $\mathcal{M}^{mnpr}$ are actually \emph{not} the full amplitudes; these expressions sill need to be contracted with the appropriate vertex factors $V_{ijk}$ or $V_c^{ijkl}$. The different choices of these terms yield different Witten diagrams; for example,
\begin{equation*}
\begin{aligned}
\mathcal{M}_{4s}=&V^{12m}V^{34n}\mathcal{M}^{mn}\;, &
\mathcal{M}_{5a}=&V^{12m}V^{3np}V^{45r}\mathcal{M}^{mnpr}\\
\mathcal{M}_{5b}=&V^{12m}V^{345n}_c\mathcal{M}^{mn}\;,\hspace{-0.2in} &
\mathcal{M}_{6b}=&V^{12m}V^{3np}V^{456r}_c\mathcal{M}^{mnpr}
\end{aligned}
\end{equation*}
which match the respective amplitudes calculated by brute force in \cite{Albayrak:2018tam}.
\section{Discussion and Future Directions}
\label{sec:conclusion}

In this paper we discussed AdS amplitudes for gauge fields and developed a formalism that considerably simplifies the calculation of any tree level Witten diagram. This formalism is based on two observations: first, the calculation of truncated amplitudes reduce to computations of scalar graphs in a judiciously chosen basis, and second, the amplitudes for the scalar graphs can be extracted by mere algebraic means. With these observations, we can obtain any tree level amplitude for gauge fields in a systematic and elegant fashion.

The advantage of our procedure is twofold. Working in the appropriate basis,
we can relate several calculations to each other, drastically simplifying the overall complexity. Indeed, in the conventional approach, the number of integrals required for calculation increases exponentially.\footnote{For example, one needs to calculate four different integrals for a five-point Witten diagram, see \equref{3.23-3.24} in \cite{Albayrak:2018tam}. In our formalism, we only did one explicit calculation, \equref{\ref{eq: five point straight amplitude}}, and obtained the rest trivially by \equref{\ref{eq: five point amplitudes}}.} The other advantage of our technique is that it is purely algebraic, which allows us to bypass the bulk integrations altogether. However, it is an open question how to extend this formalism beyond the gauge boson and to other dimensions.\footnote{The algorithm we used to compute scalar graphs is applicable only for AdS$_4$. It is intriguing to know whether analogous algorithms exist in other dimensions.}

We hope that our formalism can be utilized to generate more data points in the study of amplitudes in Anti-de Sitter space. Knowledge of higher point amplitudes may result in unraveling deeper physical and mathematical insights, similar to what the flat space program has achieved over the last decade. In this sense, we see our work as a complementary approach to those developments; for instance, it would be interesting to explore a possible connection between AdS amplitudes and geometric structures like the amplituhedron.

\begin{acknowledgments}
We would like to thank Paolo Benincasa, David Meltzer, and Sarthak Parikh for discussions. We also would like to thank Nima Arkani-Hamed  and Paolo Benincasa for sharing the draft of their unpublished paper which tackles related problems. CC thanks R. Loganayagam and Suvrat Raju for discussions and guidance. All the figures were created in Tikz. SK was supported by DRFC Discretionary Funds from Williams College.
 SA is supported by NSF grant PHY-1350180 and Simons Foundation grant 488651.
\end{acknowledgments}

\appendix
\section{A bestiary for star triangle topology}
\begin{figure}
	\centering
	\includegraphics[width=4.5cm]{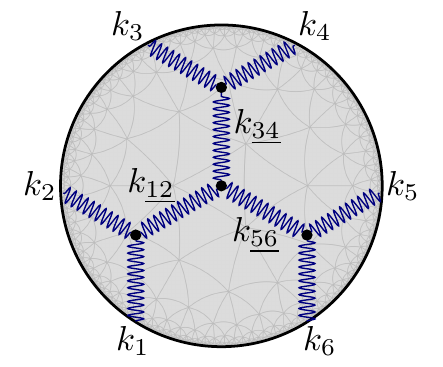}
	\caption{\label{fig:6pt} The six point diagram out of star-triangle topology}
\end{figure}

In this appendix, we will make a detailed analysis of \figref{\ref{fig:6pt}}. The truncated amplitude can be written as 
\begin{equation}
\label{eq: 6pt truncated}
\begin{aligned}
\includegraphics[width=2.45cm]{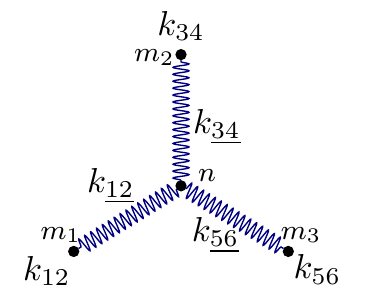}
\end{aligned}
=\sum\limits_{i,j,k=1}^{2}
\Pi^{(i)m_1n_1}_{k_{\underline{12}}}
\Pi^{(j)m_2n_2}_{k_{\underline{34}}}
\Pi^{(k)m_3n_3}_{k_{\underline{56}}}
\mathcal{A}^{(ijk)}_6
\end{equation}
where we define straight-only graph $\mathcal{A}^{(111)}_6$ as
\begin{equation}
\mathcal{A}^{(111)}_6\coloneqq
\begin{aligned}
\includegraphics[width=2.2cm]{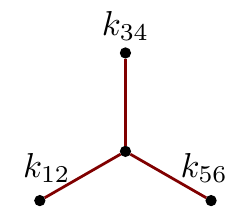}
\end{aligned}
\end{equation}
The other pieces such as $\mathcal{A}^{(112)}$ or $\mathcal{A}^{(121)}$ are same as $\mathcal{A}^{(111)}$ except the replacement of the straight leg with the crossed one, e.g.
\begin{equation}
\hskip -.08in
\mathcal{A}^{(211)}_6\coloneqq
\begin{aligned}
\includegraphics[width=2.1cm]{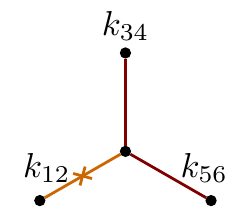}
\end{aligned}\;,\;\;
\mathcal{A}^{(222)}_6\coloneqq
\begin{aligned}
\includegraphics[width=2.1cm]{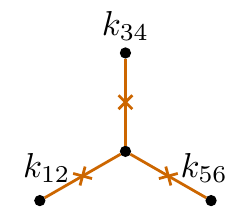}
\end{aligned}
\end{equation}
We can relate all $\mathcal{A}^{(ijk)}$ to $\mathcal{A}^{(111)}$ by taking appropriate limits:
\begin{equation}
\label{eq: M6 for crossed lines}
\mathcal{A}^{(211)}_6=\lim\limits_{k_{\underline{12}}\rightarrow 0}\mathcal{A}^{(111)}_6 \;,\;\;
\mathcal{A}^{(221)}_6=\lim\limits_{\substack{k_{\underline{12}}\rightarrow 0\\k_{\underline{34}}\rightarrow 0}}\mathcal{A}^{(111)}_6\;,\;\;\dots
\end{equation}
This reduces the whole calculation to that of the straight-only graph. We will carry out that computation using our algorithm as an explicit demonstration.

Our diagram satisfies a neat symmetry hence it is sufficient to calculate only one subgraph. More explicitly, at the top layer, we have the decomposition
\begin{equation}
\hskip -0.1in
\begin{aligned}
\includegraphics[width=1.94cm]{star1}
\end{aligned}
=
\begin{aligned}
\includegraphics[width=1.94cm]{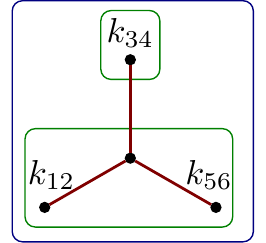}
\end{aligned}
+	\begin{aligned}
\includegraphics[width=1.94cm]{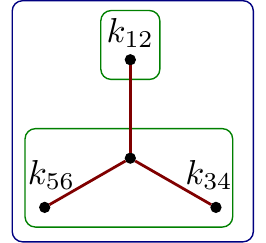}
\end{aligned}
+	\begin{aligned}
\includegraphics[width=1.94cm]{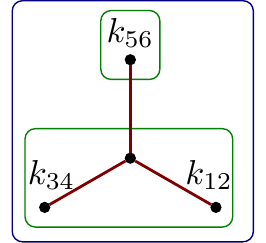}
\end{aligned}
\end{equation}
where each term is related to one another by permutation of $k_{12},k_{34},$ and $k_{56}$.

As we go further, subgraphs in deeper layers also satisfy similar symmetries. For example, the decomposition of the second graph above can be written as
\begin{equation}
\label{eq:starDecomposition}
\begin{aligned}
\includegraphics[width=2.6cm]{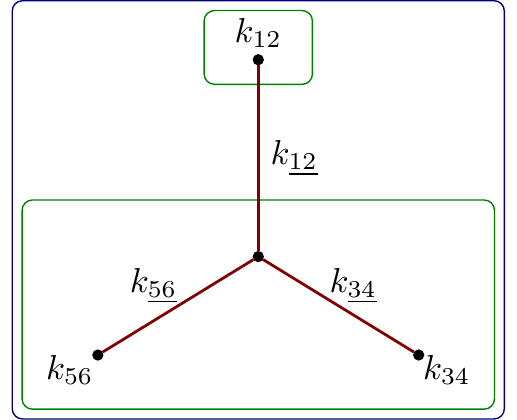}
\end{aligned}
= 
\begin{aligned}
\includegraphics[width=2.6cm]{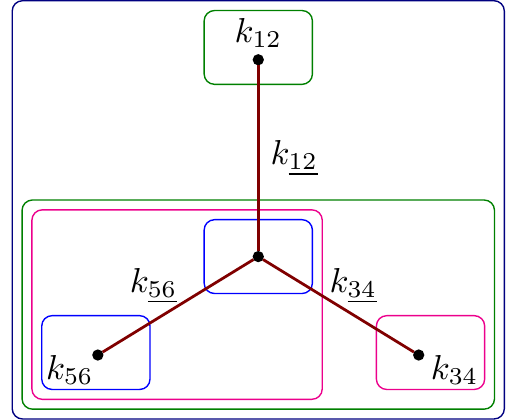}
\end{aligned}
+
\begin{aligned}
\includegraphics[width=2.6cm]{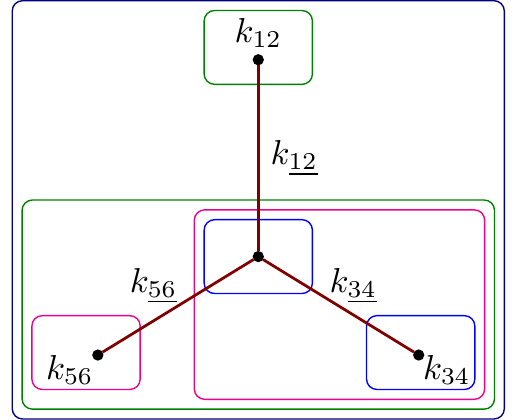}
\end{aligned}
\end{equation}
Clearly, we can obtain the second diagram by the replacement $\{k_{34},k_{\underline{34}}\} \leftrightarrow \{k_{56},k_{\underline{56}}\}$, hence we get the full result for the scalar truncated amplitude as
\begin{equation}
\label{eq: star triangle amplitude}
\mathcal{A}^{(111)}_6=\Big(\mathcal{I}\,+\, 34\leftrightarrow 56\Big)+
\left(\scriptsize\begin{aligned}
12\rightarrow 34\\
34\rightarrow 56\\
56\rightarrow 12
\end{aligned}\normalsize\right)
+\left(\scriptsize\begin{aligned}
12\rightarrow 56\\
56\rightarrow 34\\
34\rightarrow 12
\end{aligned}\normalsize\right)
\end{equation}
where $ab\rightarrow cd$ stands for $\{k_{ab},k_{\underline{ab}}\} \rightarrow \{k_{cd},k_{\underline{cd}}\}$. Here, $\mathcal{I}$ denotes the first diagram in the right hand side of \equref{\ref{eq:starDecomposition}}. We can immediately read off its value by our algorithm:
\begin{multline}
\mathcal{I}=\frac{1}{k_{123456}k_{12\underline{12}}k_{\underline{12}3456}k_{34\underline{34}}k_{\underline{12}56\underline{34}}k_{56\underline{56}}k_{\underline{12}\,\underline{34}\,\underline{56}}}
\end{multline}
When we insert this expression into \equref{\ref{eq: star triangle amplitude}}, we obtain
\begin{multline}
\mathcal{A}^{(111)}_6=\frac{1}{k_{12\underline{12}}k_{34\underline{34}}k_{56\underline{56}}k_{123456}k_{\underline{12}\,\underline{34}\,\underline{56}}}
\\
\times\left(
\frac{k_{12\underline{12}34\underline{34}\,\underline{56}\,\underline{56}}}{k_{1234\underline{56}}k_{\underline{12}34\underline{56}}k_{\underline{34}12\underline{56}}}+\text{permutations}
\right)\;.
\end{multline}
We can now use \equref{\ref{eq: M6 for crossed lines}} to get all $\mathcal{A}^{(ijk)}_6$ and substitute them into \equref{\ref{eq: 6pt truncated}} to get the full truncated amplitude.

One naively sees a divergence in the calculation of the piece $\mathcal{A}^{(222)}_6$. This poses no issue, as that term does not contribute to the result since it vanishes once it is contracted with the vertex factor. This becomes transparent if we rewrite the three point vertex as a sum of three projectors, i.e.
\begin{equation}
V_{ijk}(\ve{k}_1, \ve{k}_2, \ve{k}_3)\coloneqq{}i\sqrt{2}\Big(
\ve{k}_1^l\eta_{j[i}\eta_{l]k}
+\ve{k}_2^l\eta_{k[j}\eta_{l]i}
+\ve{k}_3^l\eta_{i[k}\eta_{l]j}\Big)\;.
\end{equation}
As the crossed lines come with factors of $\ve{k}_1^i, \ve{k}_2^j$, and $\ve{k}_3^k$, the number of non-vanishing terms in the vertex factor decreases per crossed line entering the vertex; e.g., there are only two pieces for $\mathcal{A}^{(112)}$, only one piece for $\mathcal{A}^{(122)}$, and no piece for $\mathcal{A}^{(222)}$.

With the truncated star triangle diagram at hand, we can calculate several different Witten diagrams by attaching different vertex structures. The simplest such diagram is the six-point amplitude shown in \figref{\ref{fig:6pt}}.

\bibliography{savanreference}
\end{document}